\documentclass[aps,prd,superscriptaddress,groupedaddress,nofootinbib,nobibnotes]{revtex4}

\usepackage{graphicx}
\usepackage{dcolumn}
\usepackage{bm}
\usepackage{amssymb}
\usepackage{epstopdf}
\usepackage{amsmath}
\usepackage{amsfonts}
\usepackage{color}
\usepackage{mathrsfs}
\usepackage{verbatim}

\newcommand{\be}{\begin{equation}}
\newcommand{\ee}{\end{equation}}
\newcommand{\ba}{\begin{eqnarray}}
\newcommand{\ea}{\end{eqnarray}}
\newcommand{\nn}{\nonumber}
\newcommand{\barr}{\begin{array}}
\newcommand{\earr}{\end{array}}

\newcommand\lsim{\mathrel{\rlap{\lower4pt\hbox{\hskip1pt$\sim$}}
        \raise1pt\hbox{$<$}}}
\newcommand\gsim{\mathrel{\rlap{\lower4pt\hbox{\hskip1pt$\sim$}}
        \raise1pt\hbox{$>$}}}

\def\fnl{f_{NL}}
\def\gnl{g_{NL}}
\def\tnl{\tau_{NL}}
\def\Corr{\mbox{Corr}}
\def\Mmin{M_{\rm min}}
\def\x{{\bf x}}
\def\k{{\bf k}}

\def\neff{n_{\rm eff}}
\def\kmin{k_{\rm min}}
\def\kmax{k_{\rm max}}
\def\keq{k_{\rm eq}}
\def\sinc{{\rm sinc}}
\def\Nz{N_{\rm bins}}
\def\Lz{L_\parallel}
\def\z{\chi}
\def\lz{l_\chi}
\def\n1{n_0}

\def\dh{{\bm \delta}_h}
\def\b{{\bm b}}
\def\C{{\bm C}}
\def\P{{\bm P}}
\def\E{{\bm E}}
\def\ep{{\bm \epsilon}}

\begin{document}

\title{Using Large Scale Structure to test Multifield Inflation}

\author{Simone Ferraro}
\affiliation{Princeton University Department of Astrophysical Sciences, Princeton NJ 08544, USA}

\author{Kendrick M.~Smith}
\affiliation{Perimeter Institute for Theoretical Physics, Waterloo ON N2L 2Y5, Canada}

\date{\today}

\begin{abstract}
Primordial non-Gaussianity of local type is known to produce a scale-dependent contribution to the galaxy bias.
Several classes of multi-field inflationary models predict non-Gaussian bias which is \textit{stochastic}, 
in the sense that dark matter and halos don't trace each other perfectly on large scales. 
In this work, we forecast the ability of next-generation Large Scale Structure surveys to constrain common types of primordial non-Gaussianity 
like $\fnl$, $\gnl$ and $\tnl$ using halo bias, including stochastic contributions. 
We provide fitting functions for statistical errors on these parameters
which can be used for rapid forecasting or survey optimization.
A next-generation survey with volume $V = 25 h^{-3}$Gpc$^3$, median redshift $z = 0.7$ and mean bias $b_g = 2.5$, can achieve $\sigma(\fnl) = 6$, $\sigma(\gnl) = 10^5$ and $\sigma(\tnl) = 10^3$ if no mass information is available.
If halo masses are available, we show that optimally weighting the halo field in order to reduce sample variance can achieve $\sigma(\fnl) = 1.5$, $\sigma(\gnl) = 10^4$ and $\sigma(\tnl) = 100$ if halos with mass down to $\Mmin = 10^{11}$ $h^{-1} M_\odot $ are resolved, outperforming Planck by a factor of 4 on $\fnl$ and nearly an order of magnitude on $\gnl$ and $\tnl$.
Finally, we study the effect of photometric redshift errors and discuss degeneracies between different non-Gaussian parameters,
as well as the impact of marginalizing Gaussian bias and shot noise.
\end{abstract}

\maketitle

\section{Introduction}

The study of the statistical properties of the primordial fluctuations beyond the power spectrum has enormous constraining power on inflationary models. 
While single field slow roll inflation predicts Gaussian fluctuations \cite{Maldacena:2002vr, Creminelli:2004yq}, for which all of the information lies in the primordial power spectrum, 
a wealth of alternative models (in particular multifield models) can produce detectable non-Gaussianity.

At the time of writing the best constraints come from measurements of the \textit{Cosmic Microwave Background} radiation (CMB) \cite{wmap9, planck}. However these measurements are already close to being cosmic-variance limited since the CMB is produced on a two dimensional surface, and small scales are suppressed
by Silk damping (although future measurements of $E$-mode polarization may improve statistical errors by a factor $\approx \sqrt{2}$).

With the ability of extracting 3D information and smaller scale modes, Large Scale Structure (LSS) has the potential of soon reaching and improving CMB constraints. The simplest forms of primordial local non-Gaussianity have been shown to leave a very distinctive imprint in the halo power spectrum, in the form of a scale-dependent bias proportional to $k^{-2}$ \cite{Dalal:2007cu, Matarrese:2008nc}. This has been recently generalized \cite{Desjacques:2011mq, stochastic, equivalence} to arbitrary inflationary models.  In some multifield models, non-Gaussian halo bias can be {\em stochastic}: the halo and matter fields are not 100\% correlated on large scales \cite{Tseliakhovich:2010kf, stochastic, Smith:2010gx}.  This is an important observational signature which can be used to discriminate between models which do and do not predict stochastic bias.

Analysis of existing LSS datasets yield constraints that are comparable to the ones from WMAP \cite{Ross2012, Leistedt2014, Karagiannis2014, GiannantonioNGb, Slosar:2008hx, Xia:2010pe, Xia:2010yu, Ho:2013lda}, with almost all of them being limited by spurious large-scale power due to systematics (extinction, stellar contamination, imperfect calibration, etc. \cite{Pullen2012, Huterer:2012zs}). Recently developed techniques such as mode projection and extensions \cite{Pullen2012, Leistedt2014, Leistedt:2013gfa, Leistedt:2014wia} or weights method \cite{Ross2012, Karagiannis2014} are very promising ways to reduce the impact of systematics.

The $k^{-2}$ scaling makes the signal largest on the very largest scales, which are affected by cosmic variance.
In \cite{Seljak_SV, Hamaus, GilMarin}, it was observed that cosmic variance may be partially cancelled by splitting the sample in bins of different halo mass,
and taking a linear combination of halo fields such that the Gaussian bias terms $(b_g \delta_m)$ nearly cancel, but
non-Gaussian bias terms of the form $(b_{NG} \delta_m / k^2)$ do not cancel.
A related idea for reducing statistical errors, proposed in \cite{SeljakPoisson, HamausPoisson}, is to reduce {\em Poisson} variance by taking a different linear combination of mass bins (essentially mass weighting) whose Poisson variance is lower than the naive $(1/n)$ expectation due to mass conservation.

Previous work \cite{Carbone, GiannantonioNG, Cunha, Hamaus, Biagetti, Ferramacho, Roth, Sefusatti:2012ye} has used the Fisher matrix formalism to forecast constraints on primordial non-Gaussianity through halo bias. 
Here we revisit the Fisher matrix calculation and provide analytically motivated fitting functions 
that are intended to be convenient for rapid forecasting or survey optimization.
We study some issues which are observationally relevant like the impact of marginalizing Gaussian bias and shot noise, and the impact of photometric redshift errors. We then extend the multi-tracer method of \cite{Hamaus, Biagetti, Ferramacho, Yamauchi:2014ioa} to include the effects of stochastic bias and to distinguish $\fnl$ from $\gnl$, which are completely degenerate when only a single tracer population is available.
Finally, we discuss separating the non-Gaussian parameters $\fnl$, $\gnl$, and $\tnl$, clarifying results in the literature and giving quantitative forecasts. 

The paper is organized as follows:  In Section \ref{sec:definitions} we introduce our notation and formalism, as well as discuss possible consequences of the recent claims of the BICEP2 collaboration about the amplitude of primordial tensor modes. The single-tracer case is treated analytically and numerically in detail in Section \ref{sec:single_bin}, while in Section \ref{sec:general_considerations}, we discuss the effect of marginalization and redshift errors on our forecasts. In Section \ref{sec:multi_bin} we show how constraints can be improved by using mass information. The (partial) degeneracy between models is discussed in Section \ref{sec:separating_fgt}, followed by discussion and conclusions in Section \ref{sec:discussion}.

\section{Definitions and notation}
\label{sec:definitions}
\subsection{Primordial non-Gaussianity and Large Scale Structure}
The statistical properties of the primordial potential $\Phi(\k) = (3/5) \zeta(\k)$ can be completely characterized by its $N$-point connected correlation function, which we will denote by $\xi^{(N)}_\Phi$:
\be
\langle \Phi({{\k}_1}) \Phi({{\k}_2}) \cdots  \Phi({{\k}_N})\rangle_{\rm c} 
  = (2\pi)^3 \delta_{\rm D}({\k}_1 +  {\k}_2 +  \dots + {\k}_N) \, \xi_\Phi^{(N)}({\k}_1, {\k}_2, \dots, {\k}_N)\ , \label{equ:xi}
\ee
It is customary to define the potential power spectrum $P_{\Phi}(k) = \xi^{(2)}_{\Phi}(\k, -\k)$ and 
the dimensionless power spectrum $\Delta^2_{\Phi}(k) = k^3 P_{\Phi}(k) / 2 \pi^2$.

We shall consider a model with primordial bispectrum and trispectrum parametrized by two parameters $\fnl$ and $\tnl$, which here we will assume to be independent\footnote{It can be shown on general grounds that they have to satisfy the Suyama-Yamaguchi \cite{SY, Smith:2011if} inequality $\tnl \geq (\frac{6}{5} \fnl)^2$. Specific theories of inflation will predict particular relations between $\fnl$ and $\tnl$.}
\ba \xi^{(3)}_{\Phi}(k_1, k_2, k_3) 
&=& \fnl \big[P_{\Phi}(k_1) P_{\Phi}(k_2) + \mathrm{5 \, perms.} \big]  , \label{equ:tnl3}\\
\xi^{(4)}_{\Phi}(k_1, k_2, k_3, k_4)
&=& 2\left (\tfrac{5}{6} \right )^2 \tnl \big[P_{\Phi}(k_1) P_{\Phi}(k_2) P_{\Phi}({|{\k}_1 + {\k}_3|}) + \mathrm{23 \, perms.} \big]  \ , \label{equ:tnl4}
\ea
This can be realized for example in the curvaton model \cite{curvaton, stochastic, Tseliakhovich:2010kf}, in which the non-Gaussian gravitational potential $\Phi$ is expressed in terms of two uncorrelated Gaussian fields $\phi$ and $\psi$, with power spectra that are proportional to each other
\be
\Phi(\x) = \phi(\x) + \psi(\x) + \fnl(1+\Pi)^2 \, \left(\psi^2(\x) - \langle \psi^2 \rangle \right)\ ,  \label{eqn:taunlmodel}
\ee
where $\fnl$ and $\Pi = P_\phi(k) / P_\psi(k)$ are free parameters. In this case, we can check that $\tnl = (\frac{6}{5} \fnl)^2 (1+\Pi)$,
so that $\fnl$ and $\tnl$ are independent parameters.

The matter overdensity $\delta_m(\k,z)$ is related to the primordial potential $\Phi(\k)$ through the Poisson equation,
\be
\delta_m(\k,z) = \alpha(k,z) \Phi(\k) \ .
\ee
Here we have defined $\alpha(k,z)$ by
\be
\alpha(k,z) = \frac{2 k^2 T(k)}{3 \Omega_m H_0^2} D(z)
\ee
where $D(z)$ is the linear growth function normalized so that $D(z) = 1 / (1+z)$ in matter domination (so that 
$D(z) \approx 0.76$ at $z=0$) and $T(k)$ is the transfer function normalized to 1 at low $k$.

It can be shown that in presence of non-zero $\fnl$ or $\tnl$, the halo matter and halo-halo power spectra acquire a scale dependent bias on large scales \cite{Desjacques:2011mq, stochastic, Tseliakhovich:2010kf}:
\ba
P_{mh}(k, z) &=& \left( b_g + \fnl \frac{\beta_f}{\alpha(k,z)} \right) P_{mm}(k,z) \\
P_{hh}(k, z) &=& \left( b_g^2 + 2 b_g \fnl \frac{\beta_f}{\alpha(k,z)} + \frac{25}{36} \tnl \frac{\beta_f^2}{\alpha(k,z)^2} \right) P_{mm}(k,z) + \frac{1}{\neff}   \label{eq:ng_bias1}
\ea
Here, $b_g$ is the Eulerian halo bias, and $\beta_f$ is a non-Gaussian bias parameter which can be expressed exactly as
a derivative of the tracer density $n$ with respect to the power spectrum amplitude: $\beta_f =2 \partial \ln n / \partial \ln \Delta_{\Phi} $.
Throughout this paper, we will use the alternate expression $\beta_f = 2 \delta_c (b_g-1)$,
which is exact in a barrier crossing model with barrier height $\delta_c$ and is a good ($\approx 10$\% accurate)
fit to $N$-body simulations.
We will take $\delta_c = 1.42$, as appropriate for the Sheth-Tormen \cite{ShethTormen} halo mass function.  The $1/\neff$ term enters as a Poisson shot noise term in $P_{hh}$ due to the discrete nature of tracers. The value of $\neff$ is only approximately equal to the number density of tracers $n$ and marginalization over a constant contribution to $P_{hh}$ will be discussed in Section \ref{sec:single_bin}.

We note that if $\tnl > (\frac{6}{5} \fnl)^2$, then Eq.~(\ref{eq:ng_bias1}) implies that halo and matter fields are not 100\% correlated on large scales
even in the absence of shot noise.  This phenomenon is known as `\textit{stochastic bias}'.

Another model that we will study is one that is cubic in the potential:
\be
\Phi(\x) = \phi(\x) + \gnl \left(\phi^3(\x) - 3 \langle \phi^2 \rangle \phi(\x) \right)\  . \label{equ:gnl_model}
\ee
Here it is easy to show \cite{Smith:2011ub, Desjacques:2011mq, stochastic, equivalence} that for low $k$:
\ba
P_{mh}(k, z) &=& \left( b_g + \gnl \frac{\beta_g}{\alpha(k, z)} \right) P_{mm}(k, z)\ ,  \\
P_{hh}(k, z) &=& \left( b_g + \gnl \frac{\beta_g}{\alpha(k, z)} \right)^2 P_{mm}(k, z) + \frac{1}{\neff} \ , 
\ea
where $\beta_g = 3 \partial \ln n / \partial \fnl$ is the derivative of the tracer density with respect to $\fnl$.
In this case, the barrier crossing model prediction for $\beta_g$ does not agree well with $N$-body simulations,
and for numerical work we use fitting functions for $\beta_g$ from Section 5.3 of \cite{Smith:2011ub}.

Currently the best limits on $\fnl$ and $\tnl$ are from the Planck satellite \cite{planck}, which constrains (local) $\fnl = 2.7 \pm 5.8$ and $\tnl < 2800$ (95\% CL). Regarding $\gnl$, an independent analysis of WMAP9 data has found $\gnl = (-3.3 \pm 2.2) \times 10^5$ \cite{Planckgnl}, while the Planck Fisher matrix forecast is $\sigma(\gnl) = 6.7 \times 10^4$ \cite{Planckgnl}.
 
Throughout the paper we will assume a flat $\Lambda$CDM model as our fiducial cosmology with parameters from the Planck (2013) data release: $\Omega_m h^2$ = 0.14, $\Omega_\Lambda = 0.69$, $h = 0.68$, $\ln(10^{10} A_s) = 3.09$, $\tau = 0.09$ and $n_s = 0.96$.

\subsection{Fisher Matrix analysis}

The Fisher information matrix for a multivariate random variable $\pi$ which depends on parameters $\{\theta_\alpha\} = \{\fnl, \tnl, \gnl\}$ 
through the conditional likelihood $\mathcal{L}(\pi | \theta_\alpha)$ is given by 
\be
F_{\alpha\beta} = - \left \langle \frac{\partial^2 \ln \mathcal{L}(\pi | \theta)}{\partial \theta_\alpha \ \partial \theta_\beta} \right \rangle  \label{eq:F_general}
\ee
where the expectation value is taken over random realizations of $\pi$ for a fixed fiducial set of parameters $\theta_\alpha$.

We specialize Eq.~(\ref{eq:F_general}) to the case where $\pi = (\delta_1(\k), \cdots, \delta_N(\k))$ represents all $k$-modes of a set of Gaussian fields $\delta_i$,
and the $N$-by-$N$ covariance matrix $C_{ij}(k) = P_{\delta_i\delta_j}(k)$ depends on the parameters $\theta_\alpha$.
In this case, we have:
\be
\log \mathcal{L}(\delta_i | \theta_\alpha) = \sum_{\k} \left( -\frac{1}{2} {\rm Tr} \log C(k) - \frac{1}{2} \delta_i(\k) C^{-1}_{ij}(k) \delta_j(\k) \right)
\ee
which leads to the Fisher matrix:
\be
F_{\alpha\beta} = \sum_{\k} \frac{1}{2} {\rm Tr} \left[C^{-1} \ \frac{\partial C}{\partial \theta_\alpha} \ C^{-1} \ \frac{\partial C}{\partial \theta_\beta} \right]  \label{eq:F_gaussian}
\ee
and every term is evaluated around the fiducial cosmology (usually $\fnl = \tnl = \gnl = 0$). The (marginalized) error on $\theta_\alpha$ is given by $\sigma_\alpha = (F^{-1})^{1/2}_{\alpha\alpha}$ (no sum), while the error on $\theta_\alpha$ fixing all other parameters to their fiducial values is $\sigma_\alpha = (F_{\alpha\alpha})^{-1/2}$ (again no sum). Similarly, the covariances are given by ${\rm Cov}(\hat{\theta}_\alpha, \hat{\theta}_\beta) = (F^{-1})_{\alpha\beta}$. 

For a 3D Large Scale Structure survey with volume $V$, we replace the mode sum $\sum_{\k}$ by:
\be
\sum_{\k} \longrightarrow V \int \frac{d^3\k}{(2\pi)^3} = V \int_{\kmin}^{\kmax} \frac{dk \ k^2}{2 \pi^2}
\ee
where $\kmin = 2 \pi / V^{1/3}$ is the fundamental mode and $\kmax$ will be specified in context.

In the single-tracer case where the random variable is the halo overdensity $\delta_h$,
the fiducial covariance is the 1-by-1 matrix $C(k) = b_g^2 P_{mm}(k) + 1/n$ and the derivative terms are
(assuming that $1/\neff$ is approximately independent of the non-gaussian parameters):
\be
\frac{\partial C}{\partial \fnl} = 2 b_g \frac{\beta_f}{\alpha(k, z)} P_{mm} \ ,\hskip0.5cm \frac{\partial C}{\partial \tnl} =  \left( \frac{5}{6} \right)^2 \frac{\beta_f^2}{\alpha(k, z)^2} P_{mm}  \ , \hskip0.5cm \frac{\partial C}{\partial \gnl} = 2 b_g \frac{\beta_g}{\alpha(k, z)} P_{mm}
\ee

\subsection{A comment on tensor modes}

Recent advances in sensitivity of CMB polarization experiments have allowed the detection of $B$-modes at degree angular scale by the BICEP2 collaboration \cite{BICEP2}.  If the amplitude of the signal is entirely attributed to primordial tensor modes\footnote{At the time of writing, it is unclear what fraction of the signal is due to galactic foregrounds~\cite{Mortonson:2014bja,Flauger:2014qra}.}, it would correspond to a tensor-to-scalar ratio $r = 0.2^{+0.07}_{-0.05}$.
In this section, we comment on the implications of a detection of $r$ on local primordial non-Gaussianity.

For simplicity, assume that the inflaton produces Gaussian scalar curvature perturbation $\zeta_{inf} = (5/3) \phi$, and that there is a second `curvaton' field contributing to the scalar perturbations by an amount  $\zeta_{cur}$, but that is not driving inflation and is allowed to generate large non-Gaussianity. 

If the inflaton and curvaton are uncorrelated, the total scalar perturbation is $\Delta^2_{\zeta, tot} = \Delta^2_{\zeta, inf} + \Delta^2_{\zeta, cur}$. By definition of $r$ this is:
\be
\Delta^2_{\zeta, tot} = \frac{\Delta^2_t}{r} = \frac{8}{r} \left( \frac{H_I}{2 \pi} \right)^2
\ee
Here $\Delta^2_t = 8 (H_I/2 \pi)^2$ is the tensor power spectrum and $H_I$ is the Hubble parameter during inflation. The portion produced by the inflaton is
\be
\Delta^2_{\zeta, inf} = \frac{1}{2 \epsilon} \left( \frac{H_I}{2 \pi} \right)^2
\ee
where $\epsilon = - \dot{H}_I / H^2_I$ is one of the slow roll parameters. This means that the fraction of the scalar power generated by the inflaton is
\be
Q^2 \equiv \frac{\Delta^2_{\zeta, inf}}{\Delta^2_{\zeta, tot}} = \frac{r}{16 \epsilon}
\ee
Since slow-roll inflation requires $\epsilon \ll 1$, or more typically $\epsilon \sim 0.01$, a detection of  $r \sim 10^{-2}$ or larger would 
imply that $Q^2$ is not $\ll 1$, i.e.~a sizable fraction of the scalar perturbations must be produced by the inflaton (see also \cite{LythBICEP}).
Detectable non-Gaussianity is still possible in this model, but requires (modest) tuning, since the power spectra of $\zeta_{inf}$ and $\zeta_{cur}$
must be comparable.
A sharper conclusion we can draw is that
$\tnl$ cannot be close to its minimal value $(\frac{6}{5} \fnl)^2$, since
\be
\tnl = \left( \frac{6}{5} \fnl \right)^2 \frac{1}{1-Q^2}
\ee
in this model.
Rephrasing, if $r \gtrsim 10^{-2}$, an appreciable fraction of the non-Gaussian halo bias must be stochastic.

\section{Single tracer forecasts}
\label{sec:single_bin}

In this Section, we forecast $\fnl$ and $\tnl$ constraints obtained without use of multi-tracer techniques.
The survey will be characterized by $(V,z,b_g,1/n,\kmax)$, where $b_g$ represents the mean (number weighted) bias of the sample. Our model for $P_{hh}(k)$ is the following:
\be
P_{hh}(k, z) = \left( b_g^2 + 2 b_g \fnl \frac{\beta_f}{\alpha(k,z)} + \frac{25}{36} \tnl \frac{\beta_f^2}{\alpha(k,z)^2} \right) P_{mm}(k,z) + \frac{1}{n}~~~,
\ee
where we have taken the fiducial value of $\neff$ to be $n$. First of all we note that $\fnl$ and $\tnl$ are not (completely) degenerate in $P_{hh}$, since they generate a different scale dependence, so it's possible to distinguish them even with a single tracer population. We defer further discussion about correlations between parameters to Section \ref{sec:separating_fgt}.

From here we can calculate a 4-by-4 Fisher matrix whose rows correspond to the parameters $(\fnl,\tnl,b_g,1/\neff)$,
and compute statistical errors on each parameter, with various choices for which other parameters are marginalized.

\subsection{Some definitions}

Since the $\Phi$ power spectrum is nearly scale invariant, we can write $k^3 P_\Phi(k) = A_\Phi I(k)$, where $I(k) \equiv (k / k_0)^{n_s-1}$.
The dimensionless coefficient $A_\Phi$ is given in terms of the primordial curvature perturbation amplitude by $A_\Phi = (18\pi^2/25) \Delta_\zeta^2(k_0)$. For our fiducial cosmology based parameters from the Planck 2013 release, we find $A_\Phi \approx 1.56 \times 10^{-8}$, measured at $k_0 = 0.05$ Mpc$^{-1}$.

We define $\keq$, the scale of matter-radiation equality, to be $(aH)$ evaluated at $a = \Omega_r / \Omega_m$.
Numerically, $\keq \approx 0.0154 h$ Mpc$^{-1}$.

We will express our final results in terms of a comoving distance $R_0(z)$ and comoving tracer number density $\n1(z)$ defined by:
\ba
R_0(z)^2 &=& \frac{2 D(z)}{3 \Omega_m H_0^2} = \frac{\alpha(k,z)}{k^2 T(k)} \nn \\
\n1(z) &=& (A_\Phi R_0(z)^4 \keq)^{-1}
\ea
The length $R_0(z)$ is equal to the comoving Hubble length $1/(aH)$, times some $z$-dependent factors of order unity.
A survey with tracer density $n$ is sample variance limited at the Hubble scale if $(n/\n1) \gg \keq R_0 \approx 50$, and Poisson limited on
all scales if $(n/\n1) \ll 1$.
Numerically, $R_0(z) = 3214$ $h^{-1}$ Mpc and $\n1(z) = 3.87 \times 10^{-5}$ $h^3$ Mpc$^{-3}$
at $z=0.7$.

\subsection{Factoring the Fisher matrix}

\par\noindent
Let $F_{\alpha\beta}$ denote the 4-by-4 Fisher matrix with parameters $(\fnl, \tnl, b_g, 1/n)$.
In this Section, we will show that $F$ {\em and its inverse} can be factored in the form
\be
\Big( \mbox{Simple function of } \{ V,b_g,z \} \Big) \times \Big( \mbox{Complicated function of } \Big\{ \kmin,\kmax, \frac{b_g^2 n}{\n1(z)} \Big\} \Big)  \label{eq:factorization}
\ee
This simplifies attempts to find a fitting function, since we can fit the two factors separately.
Since the inverse Fisher matrix also factors, this simplification also works for bias-marginalized statistical errors.

To derive the factorization~(\ref{eq:factorization}), write the Fisher matrix as:
\be
F_{\alpha\beta} = \frac{V}{2} \int \frac{d^3\k}{(2\pi)^3} \frac{(\partial_\alpha P_{hh}(k,z)) (\partial_\beta P_{hh}(k,z))}{P_{hh}(k,z)^2}   \label{eq:fisher_matrix_unfactored}
\ee
Now rewrite the halo-halo power spectrum in the form:
\be
P_{hh}(k,z) = b_g^2 A_\Phi R_0(z)^4 \left( k T(k)^2 I(k)+ \keq \frac{\n1(z)}{b_g^2 n} \right)
\ee
and note that the parameter derivative $\partial_\alpha P_{hh}(k,z)$ can be factored as $f_\alpha(z) g_\alpha(k)$, 
where $\alpha$ denotes any of the parameters $\{ \fnl, \tnl, b_g, 1/n \}$, and the quantities $f,g$ are defined by:
\be
f_\alpha(z) = \left( \begin{array}{c}
   4 \delta_c b_g (b_g-1) A_\Phi R_0(z)^2  \\
   \frac{25}{36} 4 \delta_c^2 (b_g-1)^2 A_\Phi \\
   2 b_g A_\Phi R_0(z)^4  \\
      1
\end{array} \right)
\hspace{2cm}
g_\alpha(k) = \left( \begin{array}{c}
  k^{-1} T(k) I(k) \\
   k^{-3} I(k) \\
 k T(k)^2  I(k)\\
    1
\end{array} \right)
\ee
We plug the above expressions into the Fisher matrix~(\ref{eq:fisher_matrix_unfactored}) to obtain:
\be
F_{\alpha\beta} = \frac{V}{2} \frac{f_\alpha(z) f_\beta(z)}{b_g^4 A_\Phi^2 R_0(z)^8} F'_{\alpha\beta}  \hspace{1.5cm}
F^{-1}_{\alpha\beta} = \frac{2}{V} \frac{b_g^4 A_\Phi^2 R_0(z)^8}{f_\alpha(z) f_\beta(z)} F'^{-1}_{\alpha\beta}
\ee
where we have defined
\be
F'_{\alpha\beta} = \int \frac{d^3\k}{(2\pi)^3} \frac{g_\alpha(k) g_\beta(k)}{[ k T(k)^2 I(k) + \keq \n1(z) / (b_g^2 n)]^2}
\ee
Since $F'_{\alpha\beta}$ and its inverse only depend on $\{ \kmin,\kmax, b_g^2 n/ \n1(z) \}$, we have now derived the factorization~(\ref{eq:factorization}).

It will be convenient to specialize the above factorization to the cases $\alpha=\beta=\fnl$ and $\alpha=\beta=\tnl$.
If we do not marginalize either $b_g$ or $1/n$ (and set $\tnl = 0$ when forecasting $\fnl$ and vice versa), the statistical errors on $\fnl$ and $\tnl$ are given by:
\ba
\sigma(\fnl) &=& \frac{\sqrt{2}}{4\delta_c} \frac{b_g}{b_g-1} R_0(z)^2 V^{-1/2} (F'_{\fnl})^{-1/2} \nn \\
\sigma(\tnl) &=& \left( \frac{6}{5} \right)^2 \frac{\sqrt{2}}{4\delta_c^2} \left( \frac{b_g}{b_g-1} \right)^2 R_0(z)^4 V^{-1/2} (F'_{\tnl})^{-1/2}  \label{eq:factorization_ft}
\ea
where:
\ba
F'_{\fnl} &=& \int \frac{d^3\k}{(2\pi)^3} \left( \frac{k^{-1} T(k) I(k)}{k T(k)^2 I(k)+ \keq \n1(z) / (b_g^2 n)} \right)^2  \nn \\
F'_{\tnl} &=& \int \frac{d^3\k}{(2\pi)^3} \left( \frac{k^{-3} I(k)}{k T(k)^2 I(k) + \keq \n1(z) / (b_g^2 n)} \right)^2   \label{eq:fprime_ft}
\ea
To marginalize over $b_g$ and/or $1/n$, we would replace matrix elements of $F'$ in Eq.~(\ref{eq:factorization_ft})
by matrix elements of an appropriate inverse Fisher matrix.

\subsection{Sample Variance and Poisson limits; qualitative behavior}
\label{SV_P_analytic}

As an illustration of the factorization in the previous Section, let's derive approximate expressions
for $\sigma(\fnl), \sigma(\tnl)$ in the sample variance dominated limit $n/\n1(z) \gg \keq R_0(z)$ and Poisson dominated limit $n/\n1(z) \ll 1$, without
bias marginalization (and setting $n_s=1$ for this subsection).  First we take limits of Eq.~(\ref{eq:factorization_ft}), obtaining:
\ba
F'_{\fnl} & \rightarrow & \frac{1}{2 \pi^2} \kmin^{-1}  \hspace{4.1cm} \mbox{(sample variance dominated)} \nn \\
          & \rightarrow & \frac{Z}{2\pi^2} \keq^{-1} \left( \frac{b_g^2 n}{\n1(z)} \right)^2 \hspace{2.5cm} \mbox{(Poisson dominated)} \nn \\
F'_{\tnl} & \rightarrow & \frac{1}{10 \pi^2} \kmin^{-5} \hspace{3.9cm} \mbox{(sample variance dominated)} \nn \\
          & \rightarrow & \frac{1}{6\pi^2} \kmin^{-3} \keq^{-2} \left( \frac{b_g^2 n}{\n1(z)} \right)^2 \hspace{1.8cm} \mbox{(Poisson dominated)}
\ea
where we have defined the dimensionless number $Z = \keq^{-1} \int dk \, T(k)^2$.
Plugging into Eq.~(\ref{eq:factorization_ft}) to get parameter errors, and taking $\kmin = 2\pi/V^{1/3}$, we get the following approximate limits: 

\ba
\sigma(\fnl) & \rightarrow & 2.77 \frac{b_g}{b_g-1} \left( \frac{V}{R_0(z)^3} \right)^{-2/3} 
                   \hspace{4.9cm} \mbox{(sample variance dominated)} \nn \\ 
             & \rightarrow & 0.95 \frac{b_g}{b_g-1} (\keq R_0(z))^{1/2} \left( \frac{V}{R_0(z)^3} \right)^{-1/2} \left( \frac{b_g^2 n}{\n1(z)} \right)^{-1} 
                   \hspace{1cm} \mbox{(Poisson dominated)} \nn \\
\sigma(\tnl) & \rightarrow & 248 \left( \frac{b_g}{b_g-1} \right)^2 \left( \frac{V}{R_0(z)^3} \right)^{-4/3} 
                   \hspace{4.3cm}  \mbox{(sample variance dominated)} \nn \\
             & \rightarrow & 30.6 \left( \frac{b_g}{b_g-1} \right)^2 (\keq R_0(z)) \left( \frac{V}{R_0(z)^3} \right)^{-1}
                                 \left( \frac{b_g^2 n}{\n1(z)} \right)^{-1} \hspace{0.9cm} \mbox{(Poisson dominated)} \label{eq:sv_poisson_limits}
\ea

\begin{figure}[!ht]
\centerline{\includegraphics[width=12cm]{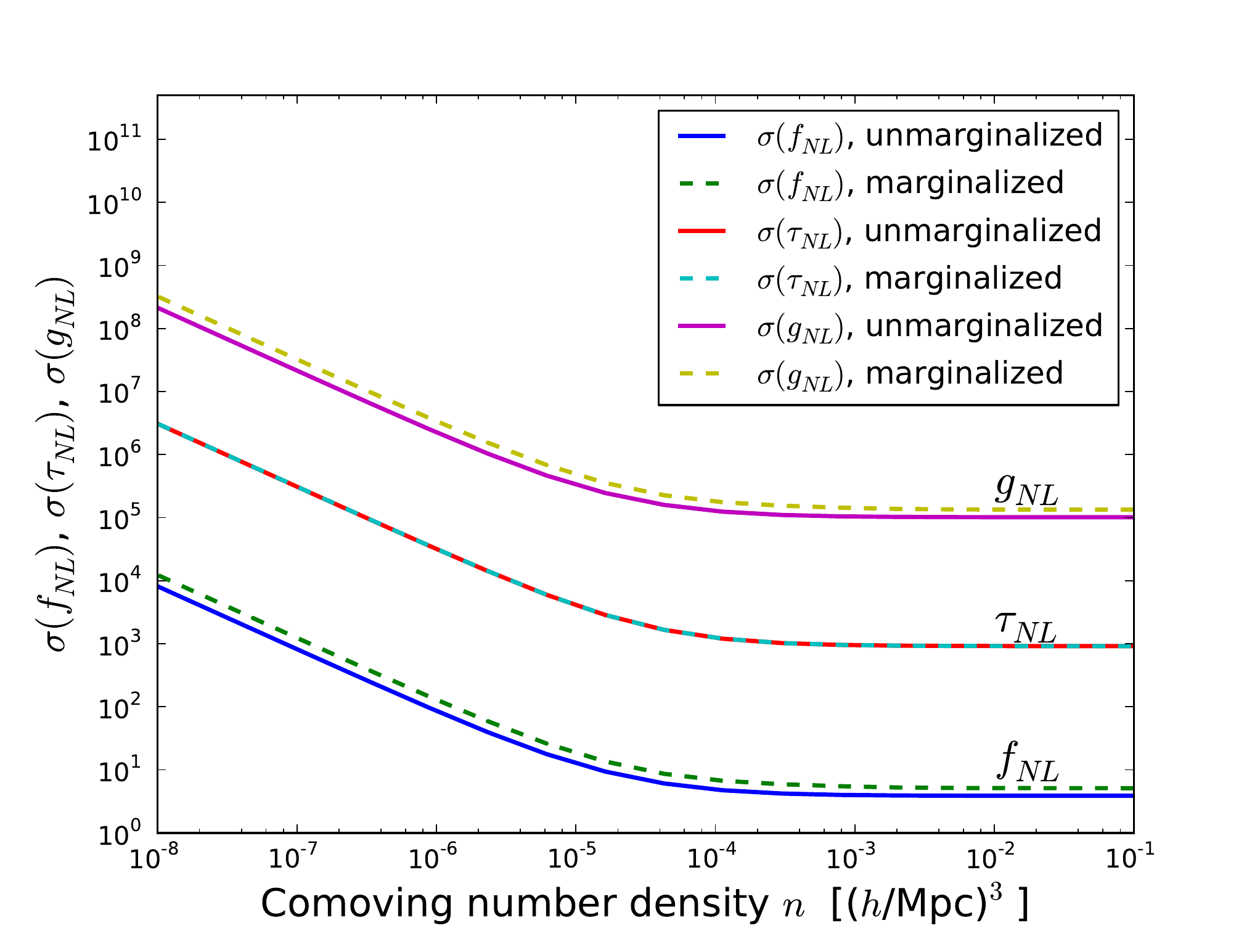}}
\caption{Statistical errors on $\fnl$ (bottom) and $\tnl$ (middle) and $\gnl$ (top) for varying tracer density $n$, for our fiducial survey with volume $V=25$ $h^{-3}$ Gpc$^3$, 
redshift $z=0.7$, tracer bias $b_g =2.5$ and maximum wavenumber $\kmax=0.1$ $h$ Mpc$^{-1}$. The `marginalized' case (dashed lines) refers to marginalization over Gaussian bias and a 20\% Gaussian prior on $1/\neff$ around the fiducial value $1/\neff = 1/n$. When forecasting each parameter $\{ \fnl, \tnl, \gnl \}$, the other two parameters are set to zero. Constraining $\gnl$ is discussed in Section \ref{sec:gnl}, while degeneracies and their covariance are discussed in Section \ref{sec:separating_fgt}.} 
\label{fig:sigmaofn}
\end{figure}

As we expected, the statistical errors are independent of tracer density $n$ in the sample variance limited case, while they scale as $1/n$ in the Poisson limit. This behavior becomes very clear in the numerical results shown in Figure \ref{fig:sigmaofn}.

We also notice the errors often depend on volume in a way which differs from the usual $V^{-1/2}$ scaling.
This happens when the $k$-integral for the Fisher matrix element diverges at low-$k$, so that most of the statistical weight comes from the survey scale $\kmin = 2\pi / V^{1/3}$. 
This divergence always occurs for $\tnl$, so the $\tnl$ constraint is always dominated by the largest-scale modes in the survey (i.e.~a few modes).
For $\fnl$ this depends on the level of Poisson noise; in the sample variance limit the statistical weight is dominated by the largest scale modes, but in
the Poisson dominated limit the statistical weight is distributed over a range of scales between $\kmin$ and $\keq$.

We also note that in the Poisson dominated case, the last line of~(\ref{eq:sv_poisson_limits}) can be rewritten: 
\be
\sigma(\tnl) = 30.6 \left( \frac{1}{b_g-1} \right)^2 (\keq R_0(z)^4 \n1(z)) \frac{1}{nV}  \hspace{0.9cm} \mbox{(Poisson dominated)}
\ee
i.e.~$\sigma(\tnl)$ only depends on $n,V$ through the total number of tracers $(nV)$ in the Poisson-dominated case.

The analytic results in this subsection are approximate (we have assumed $n_s =1$ and $T(k)=1$) and shouldn't be used in forecasting. In Figure \ref{fig:sigmaofn} we show the numerical results and in the next subsection we give fitting functions which work at the few percent level and include the effect of non-trivial $n_s$ and $T(k)$.

\subsection{Fitting functions}
Motivated by the analytically discussion of the previous Section, here we present fitting functions for $\sigma(\fnl)$ and $\sigma(\tnl)$ as functions of $(V, z, b_g, n)$, while fixing all of the parameters of the background cosmology to the Planck 2013 values, as explained in Section \ref{sec:definitions}. Moreover, we take $k_{max} = 0.1 h$ Mpc$^{-1}$ throughout.

As a first step, we define the quantity 
\be
\Gamma(n,z) = \frac{b_g^2 n}{\n1(z)} = b_g^2 \left(\frac{n}{1.17 \times 10^{-5} h^3 \ {\rm Mpc}^{-3}} \right) D^2(z)
\ee
To make our fitting functions self-contained, we note that
the linear growth factor $D(z)$ is well fit by~\cite{Carroll:1991mt}:
\be
D(z) \approx \frac52 \frac{\Omega_m(z)}{1+z} \left[\Omega_m(z)^{4/7} - \Omega_{\Lambda}(z) + \left(1+\frac12 \Omega_m(z) \right) \left(1+\frac{1}{70} \Omega_{\Lambda}(z) \right) \right]^{-1}
\ee
where $\Omega_m(z)$ and $\Omega_\Lambda(z)$ are defined by
\be
\Omega_m(z) = \frac{\Omega_m (1+z)^3}{\Omega_\Lambda + \Omega_m (1+z)^3}
\hspace{1cm}
\Omega_\Lambda(z) = \frac{\Omega_\Lambda}{\Omega_\Lambda + \Omega_m (1+z)^3}
\ee
Our fitting functions for $\sigma(\fnl)$ and $\sigma(\tnl)$ will be sums of sample variance and Poisson terms as follows:
\be
\sigma(\fnl) = \sigma_S(\fnl) + \sigma_P(\fnl)
\hspace{1.5cm}
\sigma(\tnl) = \sigma_S(\tnl) + \sigma_P(\tnl)
\ee
Note these are just fitting functions, and we are making no claims about the true variance decomposing into separate contributions.
Following the analytic results of Section \ref{SV_P_analytic}, we fit the individual terms with the functional forms:
\ba
\sigma_S(\fnl) &=& \mathcal{A}_S \ D(z) \ \frac{b_g }{b_g - 1}  \left(\frac{V}{V_0} \right)^{-2/3 + \epsilon_S + \frac{1}{2} \mu_S \ln(V/V_0)} \nn \\
\sigma_P(\fnl) &=& \mathcal{A}_P \ D(z) \ \frac{b_g}{b_g -1} \  \Gamma^{-1}(n,z)  \left(\frac{V}{V_0} \right)^{-1/2 + \epsilon_P + \frac{1}{2} \mu_P \ln(V/V_0)} \nn \\
\sigma_S(\tnl) &=& \mathcal{A}'_S \ D^2(z) \left( \frac{b_g }{b_g - 1} \right)^2   \left(\frac{V}{V_0} \right)^{-4/3 + \epsilon'_S + \frac{1}{2} \mu'_S \ln(V/V_0)} \nn \\
\sigma_P(\tnl) &=& \mathcal{A}'_P \ D^2(z) \left( \frac{b_g }{b_g - 1} \right)^2  \Gamma^{-1}(n,z) \left(\frac{V}{V_0} \right)^{-1 + \epsilon'_P + \frac{1}{2} \mu'_P \ln(V/V_0)}
  \label{eq:fit_parameters}
\ea
where $V_0 = 5 h^{-3}$ Gpc$^3$ and values of the remaining parameters 
depend on whether we are marginalizing over bias or not. As in the previous discussion we will consider the two cases: (i) when no marginalization is performed, and (ii) when we marginalize over the gaussian bias $b_g$ and assume a 20\% Gaussian prior on the shot noise $1/\neff$.  Best-fit parameter values in these two cases are given by:
\ba
\mbox{for }\fnl: \ \ (\mathcal{A}_S, \epsilon_S, \mu_S, \mathcal{A}_P, \epsilon_P, \mu_P) &=& 
\left\{ 
\begin{array}{cl}
   (10.7,\, 0.096,\, -0.009,\, 33.7,\, -0.039,\, 0.012) & \mbox{if $b_g,\neff$ unmarginalized}  \\
   (15.9,\, 0.002,\, 0.005,\, 54.2,\, -0.102,\, 0.037) & \mbox{if $b_g,\neff$ marginalized}
\end{array}
\right.  \nn \\
\mbox{for }\tnl: \ \ (\mathcal{A}'_S, \epsilon'_S, \mu'_S, \mathcal{A}'_P, \epsilon'_P, \mu'_P) &=& 
\left\{
\begin{array}{cl}
   (8477,\, 0.098,\, -0.037,\, 30405,\, -0.013,\, 0.000)  & \mbox{if $b_g,\neff$ unmarginalized}  \\
   (8493,\, 0.089,\, -0.030,\, 30830,\, -0.035,\, 0.015)  & \mbox{if $b_g,\neff$ marginalized}
\end{array}
\right.
\ea
This completes the description of our fitting functions for $\sigma(\fnl)$ and $\sigma(\tnl)$.
With the above definitions, we find that our fitting functions are accurate
to better than 10\% for $0.5 \le (V \, / \, h^{-3} \,\mbox{Gpc}^3) \le 50$ and
arbitrary $(b,n)$.  (Note that $\kmax=0.1$ has been assumed throughout; we will study
the effect of varying $\kmax$ in Section~\ref{ssec:marginalizing_bn}.)

From this we read off the following:
A sample variance limited survey with comoving volume $V = 25h^{-3}$ Gpc$^3$ and $b_g=2.5$ has statistical errors $\sigma(\fnl) \approx 6$ and
$\sigma(\tnl) \approx 1000$, comparable to Planck.  
Therefore, the only way to improve statistical errors beyond Planck is to measure a larger volume or to use a multi-tracer analysis, as described later.  

\subsection{Forecasts for $\gnl$}
\label{sec:gnl}
As we have briefly mentioned in Section \ref{sec:definitions}, the large scale bias in presence of primordial $\gnl$ is approximately given by 
\be
P_{hh}(k) = \left( b_g + \gnl \frac{\beta_g}{\alpha(k)} \right)^2 P_{mm}(k) + \frac{1}{n} \ ,  \label{eq:gnl_phh}
\ee
where $\beta_g = 3 \partial \ln n / \partial \fnl$.  In \cite{Smith:2011ub} we have found a fitting function for $\beta_g$:
\be
\beta_g(\nu) \approx \kappa_3 \bigg[ -0.7 + 1.4(\nu-1)^2 + 0.6(\nu-1)^3 \bigg]
  - \frac{d\kappa_3}{d\ln\sigma^{-1}} \left( \frac{\nu-\nu^{-1}}{2} \right)~~~.
\ee
where:
\be
\nu = [\delta_c(b_g-1) + 1]^{1/2}\,, \hspace{0.5cm}
\kappa_3 = 0.000329 (1 + 0.09z) \ b_g^{-0.09}\,, \hspace{0.5cm}
\frac{d\kappa_3}{d\ln\sigma^{-1}} = -0.000061 (1 + 0.22z) \ b_g^{-0.25}
\ee
with $\delta_c = 1.42$. Comparing Equation (\ref{eq:ng_bias1}) for a `pure' $\fnl$ cosmology (i.e. one in which $\tnl = \left( \frac65 \fnl \right)^2$), with Equation (\ref{eq:gnl_phh}), we find that the effect of $\gnl$ on halo bias is the same as the effect of $\fnl = (\beta_f / \beta_g) \gnl$ and therefore they are indistinguishable with a single tracer population. 
In particular, if we want forecasts on the detectability of $\gnl$ with a single tracer population assuming $\fnl = 0$,
we just write $\sigma(\gnl) = (\beta_f/\beta_g) \sigma(\fnl) \approx (2 \delta_c (b_g-1) / \beta_g) \sigma(\fnl)$ and use results from the previous subsection. Numerical results for our fiducial survey are shown in Figure \ref{fig:sigmaofn}.

As we will show in Sections \ref{sec:multi_bin} and \ref{sec:separating_fgt}, multiple tracer populations with different mass (or equivalently Gaussian bias), can allow us to distinguish between $\fnl$ and $\gnl$, thanks to the different dependence of the scale dependent correction on the Gaussian bias $b_g$.

\section{General considerations when constraining $\fnl$ from Large Scale Structure}
\label{sec:general_considerations}

\subsection{How much do statistical errors degrade when marginalizing bias and Poisson noise?}
\label{ssec:marginalizing_bn}

When analyzing data from a real survey, the values of $b_g$ and $\neff$ must be measured together with the non-Gaussian parameters, and it is important to understand the amount of information lost in doing so.
In this Section, we quantify this by forecasting statistical errors on $\fnl$ and $\tnl$ when the parameters
$b_g$ and $(1/\neff)$ are marginalized, and discuss our results as a function of $\kmax$.

We first note that $(1/\neff)$ is only approximately equal to $(1/n)$, where $n$ is the number density of tracers.
In addition to the $(1/n)$ term expected from Poisson statistics, there are several effects
which contribute constant power on large scales: non-linear galaxy bias, halo exclusion \cite{BaldaufStochastic}, tidal tensor 
bias~\cite{BaldaufTidal, ChanTidal},  and contributions from the HOD.
Throughout this section, when we marginalize $(1/\neff)$, we assign a Gaussian prior around
the fiducial value $(1/n)$ with width equal to 20\% of the value itself.

\begin{figure}[!ht]
\centerline{
   \includegraphics[width=9.5cm]{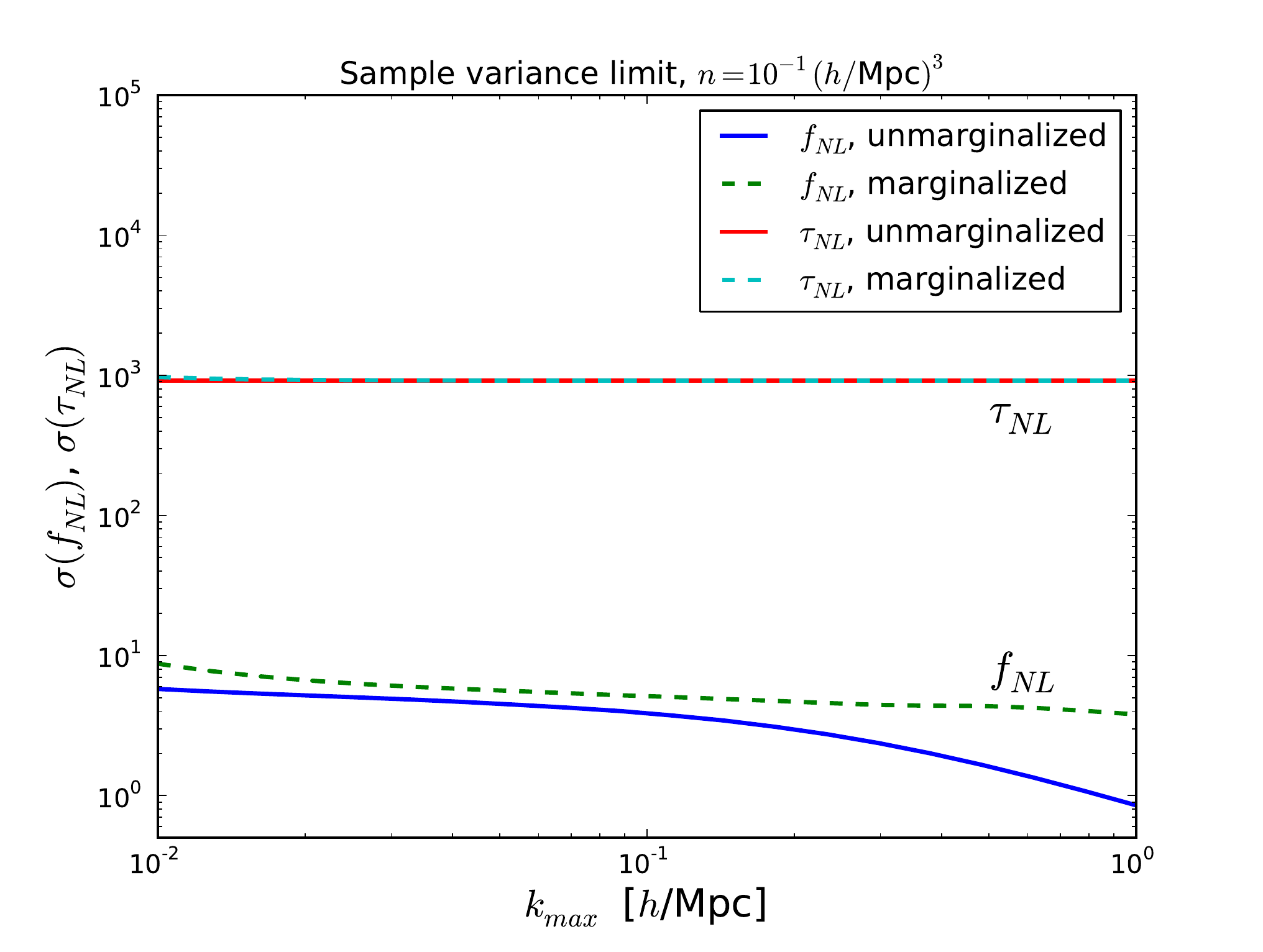}
   \includegraphics[width=9.5cm]{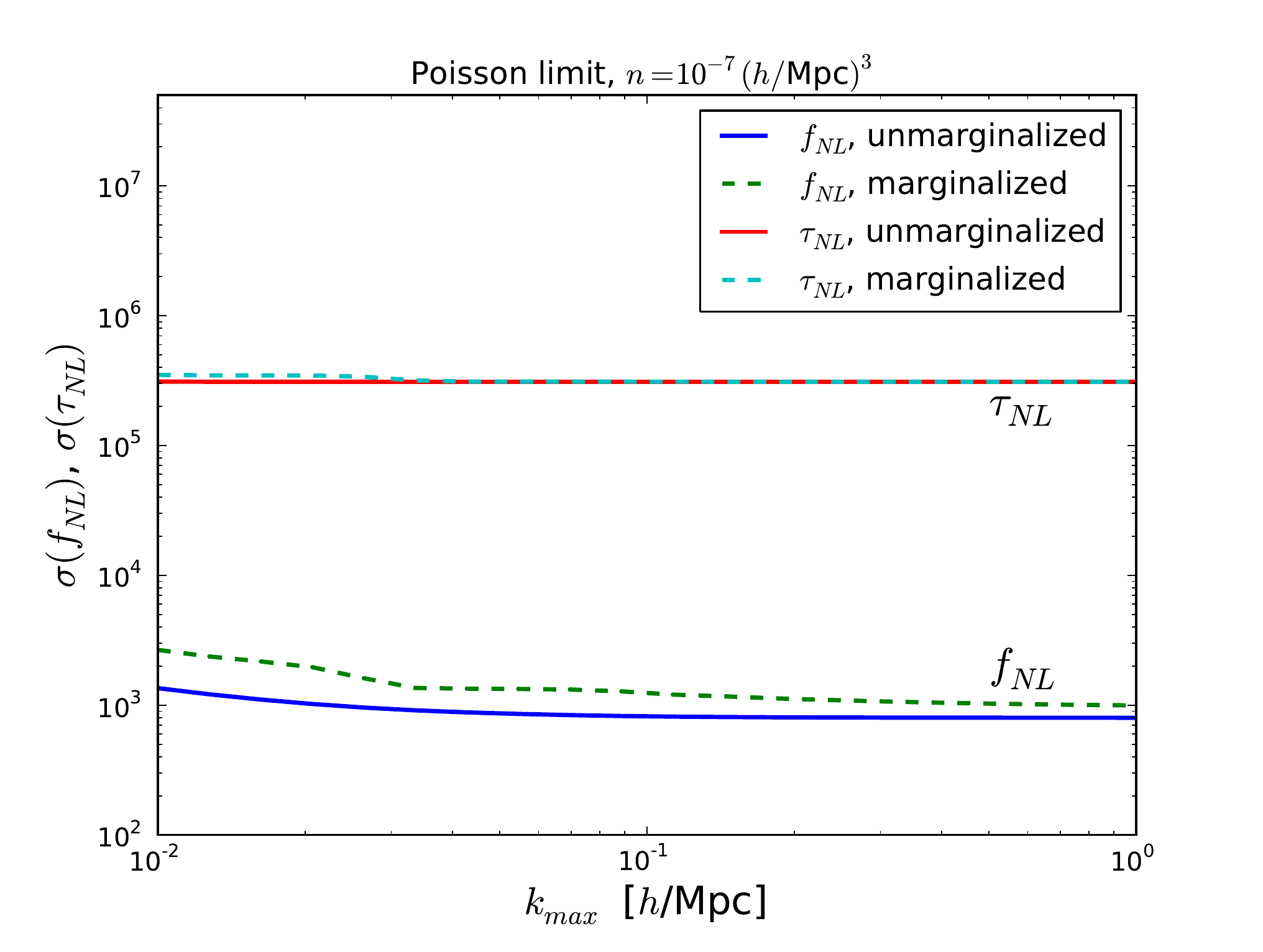}
}
\caption{Forecasts on $\fnl$ and $\tnl$ as a function of maximum wavenumber $\kmax$ in the sample variance limited (left) and Poisson limited (right) regimes. 
Here $V = 25 h^{-1}$Gpc, $z = 0.7$, and $b_g = 2.5$.}
\label{fig:kmax}
\end{figure}

In Figure~\ref{fig:kmax}, we compare statistical errors on $\fnl$ and $\tnl$ in the cases with no marginalization, or marginalization over $b_g$ and with a 20\% prior on $\neff$.
It is seen that marginalizing $b_g$ can make a large difference in $\sigma(\fnl)$, e.g.~in the sample variance limited case with $\kmax \gtrsim 0.1 h$ Mpc$^{-1}$.
This is because the non-Gaussian correction to the bias scales as $b_{NG}(k) \sim \fnl / (k^2 T(k))$, with $T(k) \sim k^{-2} \ln{(k/k_{eq})}$ for $k \gg \keq$. 
Hence, the non-Gaussian part of the bias becomes nearly degenerate with the Gaussian bias $b_g$ for $k \gg \keq$.
For $\tnl$, marginalization makes practically no difference and the statistical power increases very slowly going to higher $k$.

Based on these plots, we note that statistical errors on $\fnl$ and $\tnl$ are approximately 
saturated at $\kmax \sim 0.1 h$ Mpc$^{-1}$, if Gaussian bias is properly marginalized.
Therefore we take $\kmax = 0.1 h$ Mpc$^{-1}$ as our fiducial value in this paper.

\subsection{Redshift Errors and  3D $\rightarrow$ 2D projection}

Most observational constraints on non-Gaussianity reported in the literature have made use of projected angular correlation functions, 
rather than using redshift information. In this Section we discuss the effect of projecting three-dimensional
measurements into one or more radial bins.
This will quantify the information lost by 3D $\rightarrow$ 2D projection, and will also indicate how accurate
photometric redshifts must be in order to avoid losing information relative to an ideal 3D survey.

We use a formalism which neglects curved-sky corrections, boundary effects, and redshift evolution,
but is self-consistent given these approximations.
Consider a rectangular 3D box with periodic boundary conditions, and treat one of the three dimensions as
the `radial' direction, and the other two dimensions as `transverse'.
Let $A_\perp$ be the transverse area of the box, and let $\Lz$ be the length of
the box in the radial direction.
We divide our 3D survey in $\Nz$ radial slices and project the 3D halo field onto the closest slice. 
The case $\Nz =1$ corresponds to neglecting any redshift information (i.e. a purely 2D survey), 
while the limit $\Nz \rightarrow \infty$ corresponds to an ideal 3D survey with perfect redshifts.

Suppose that the halo field in the box is a 3D field $\delta_{3D}$ with power spectrum 
\be
P_{hh}^{3D}(k) = \left( b_g + \fnl \frac{2\delta_c(b_g-1)}{\alpha(k)} \right)^2 P_{mm}(k) + \frac{1}{n}
\ee
where the Gaussian bias $b_g$, redshift, and number density $n$ are assumed constant throughout the box.
We divide the box into $\Nz$ radial bins and project the 3D halo field into $\Nz$ two-dimensional
fields $\delta_{1}, \cdots, \delta_{\Nz}$.
We then use the 2D Fisher matrix formalism to forecast the statistical error on $\sigma(\fnl)$,
and study the dependence of $\sigma(\fnl)$ on $\Nz$.

For Fisher forecasting, we will need to compute power spectra $P_{ij}(l)$ of the 2D fields $\delta_i$.
We will avoid using the Limber approximation since we will be interested in the limit $\Nz \rightarrow \infty$
in which the Limber approximation becomes arbitrarily bad (note that we are making the flat sky approximation
throughout, but the flat sky and Limber approximations are independent).
In real space, the 3D $ \rightarrow$  2D projection is given by
\be
\delta_{i}(x,y) = \frac{\Nz}{\Lz} \int_{\z_i-\Lz/2\Nz}^{\z_i+\Lz/2\Nz} d\z \ \delta_{3D}(x,y,\z)  \label{eq:3d_to_2d_real_space}
\ee
where $(x,y)$ are transverse coordinates, $\z$ is the radial coordinate,
and $\z_i$  is the central $\z$-value of the $i$-th bin.
In Fourier space, the 3D $\rightarrow$ 2D projection is given by:
\be
\tilde\delta_i(l_x, l_y) = \int_{-\infty}^\infty \frac{d\lz}{2\pi} \ \tilde\delta_i(l_x,l_y,\lz) \ \sinc\left( \frac{\lz \Lz}{2\Nz} \right) e^{i \lz \z_i}
\ee
where $(l_x, l_y)$ is a 2D wavevector of modulus $l = (l_x^2+l_y^2)^{1/2}$ and $\sinc(x) = (\sin x)/x$.
It follows that the $\Nz$-by-$\Nz$ matrix of 2D projected power spectra is:
\be
P_{ij}(l) = \int_{-\infty}^\infty \frac{d\lz}{2\pi} \ P_{hh}^{3D} \Big( \sqrt{l^2+\lz^2} \Big) \ \sinc^2\!\left( \frac{\lz \Lz}{2\Nz} \right) 
   e^{i\lz(\z_i - \z_j)}   \label{eq:power_spectra_2d}
\ee
We will compute 2D Fisher matrices to maximum wavenumber $l_{\rm max} = 0.1$ $h$ Mpc$^{-1}$,
but take the upper limit of the $\lz$ integral in Eq.~(\ref{eq:power_spectra_2d}) large
enough that the integral converges.  Note that the 2D Fisher matrix is given by
\be
F_{\alpha \beta} = \frac{A_\perp}{2} \int \frac{d^2{\bm l}}{(2\pi)^2} {\rm Tr} \left[\P^{-1} \ \frac{\partial \P}{\partial \theta_\alpha} \P^{-1} \ \frac{\partial \P}{\partial \theta_\beta} \right]
\ee
with $\P = P_{ij}(l)$ given by Eq.~(\ref{eq:power_spectra_2d}).

\begin{figure}[!ht]
\centerline{\includegraphics[width=9.5cm]{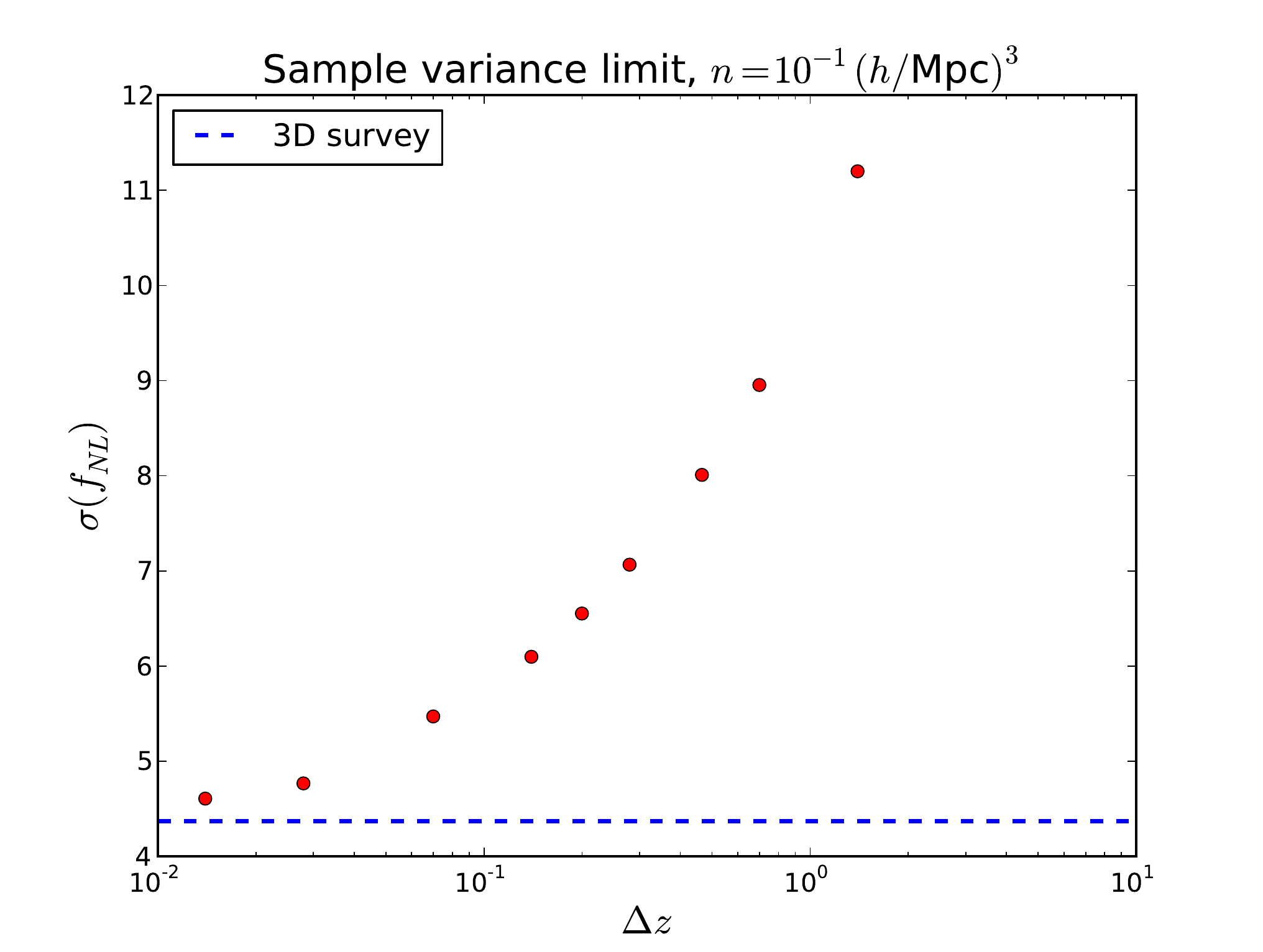} \includegraphics[width=9.5cm]{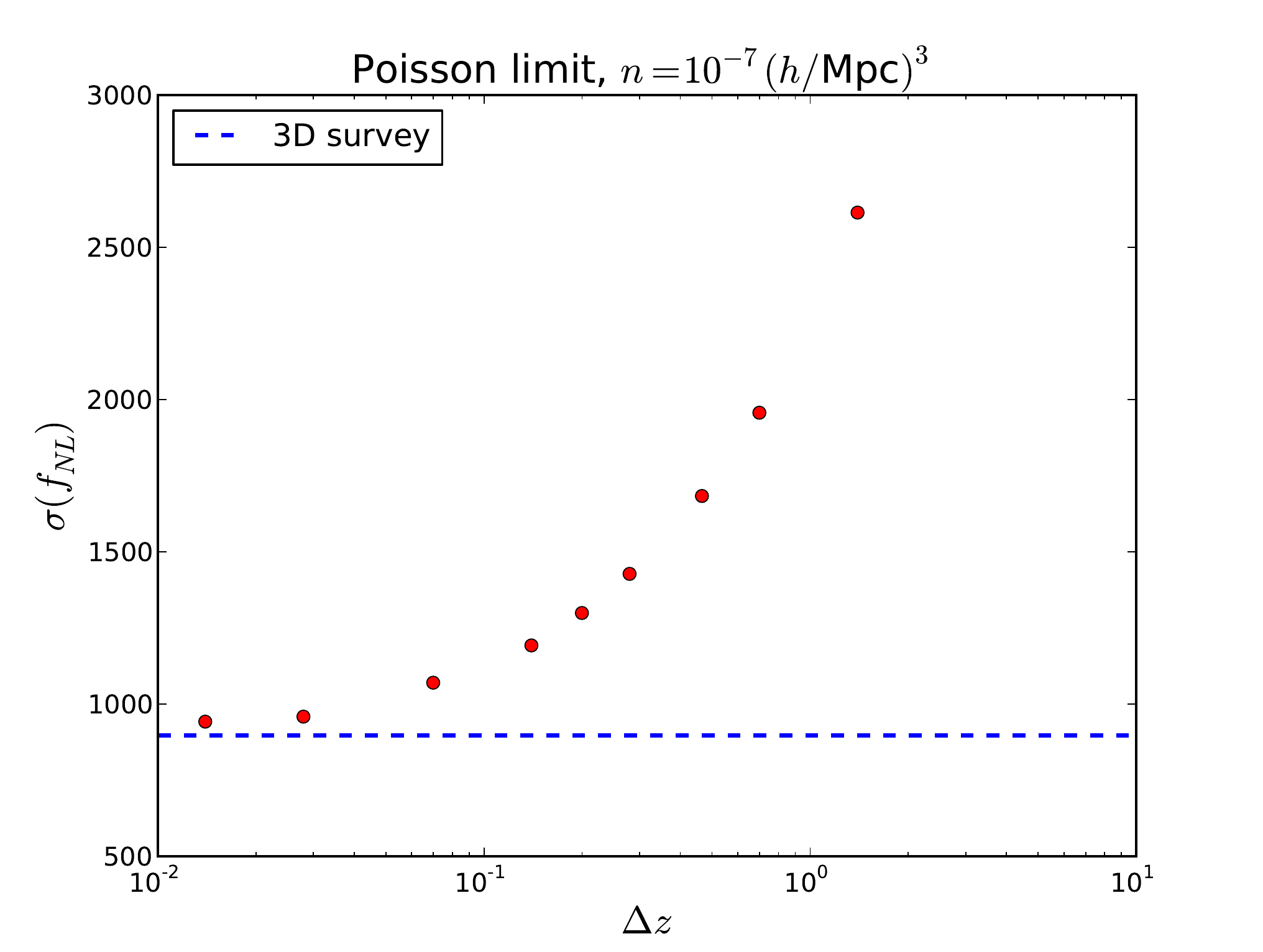}}
\caption{Dependence of statistical error $\sigma(\fnl)$ on redshift bin width $(\Delta z)$, corresponding to (from right to left) $\Nz = $ 1, 2, 3, 5, 7, 10, 20, 50, and 100.
The fiducial survey has volume $V = 25$ $h^{-3}$ Gpc$^3$, redshift $z = 0.7$ and bias $b_g=2.5$,
with a cubic geometry assumed so that $A_\perp = V^{2/3}$ and $\Lz = V^{1/3}$.
Note that the rightmost point corresponds to a 2D survey, and that the loss of information is roughly the same in the sample variance limited
and Poisson limited cases.}
\label{fig:3Dto2D}
\end{figure}

In Figure~\ref{fig:3Dto2D} we show the dependence of $\sigma(\fnl)$ on $\Nz$, in both Poisson and sample variance limited cases.
We see that completely neglecting redshift information significantly degrades the amount of information available;
the statistical error on $\fnl$ in a 2D analysis (i.e.~$\Nz=1$) is larger than the 3D case by
a factor close to 3.
However, binning in redshift bins with with redshift spread $\Delta z \sim 0.1$ or smaller
is sufficient to capture almost all of the 3D information.

We can also comment briefly on the effect of photometric redshift uncertainties.
Photometric redshifts from a multi-band instrument such as LSST are several times smaller than
$\Delta z \sim 0.1$, and therefore we expect that photometric redshift uncertainties should not 
significantly degrade statistical errors on $\fnl$. 
A caveat to this analysis is that a small fraction of {\em catastrophic} photometric redshift errors
may add large-scale power; this case should be studied separately.
(For a different approach to the study of photometric redshift errors and the closely related issue of redshift space distortions, see~\cite{Cunha,GiannantonioNG}.)

\section{Multi-tracer forecasts - Optimal Weighting}
\label{sec:multi_bin}

In this Section, we will consider multiple tracers with different Gaussian bias and show how to combine them optimally for the best constraining power on primordial non-Gaussianity.
Here we will assume that all halos above some minimum mass $\Mmin$ have been detected, and use the halo model prediction (with Sheth-Tormen
mass function) for the number density and bias.
Thus the parameters of our forecasts will be $(V,z,\Mmin)$.

Following the formalism of \cite{Hamaus}, we can divide the halo overdensity into $N \gg 1$ mass bins
$\dh = (\delta_1, \ldots, \delta_N)^T$. The number of bins will be determined by the finite mass resolution of the survey.
Assuming halos to be locally biased and stochastic tracers of the underlying density field, we can write
\be
\dh = \b \ \delta + \ep
\ee
where $\ep$ is the residual (Poisson-like) noise field, with zero mean and uncorrelated with the matter density $\delta$.
Here $b_i$ is the mean (number weighted) Gaussian bias of tracers in bin $i$:
\be
b_i = \frac{\int_{M \in {\rm bin}\ i} dM \frac{dn}{dM} b_g(M)} {\int_{M \in {\rm bin}\ i} dM \frac{dn}{dM}}
\ee
The halo covariance matrix $C_{ij}(k) = \langle \delta_i^*(\k) \delta_j(\k) \rangle$ is
\be
\C(k) = \langle \dh \dh^T \rangle = \b \b^T P_{mm}(k) + \E
\ee
where $E_{ij} = \langle \epsilon_i \epsilon_j \rangle$ is the error matrix. This has been studied analytically and with $N$-body simulations in several earlier papers (see for example \cite{Hamaus, HamausPoisson}). They find that $\E$ is approximately scale independent on the range of $k$ considered, and that the dependence on $\fnl$ is pretty weak and will be neglected here.

We will use the halo model prediction for $\E$ at low $k$, which has been shown to be a pretty good approximation to $N$-body simulations \cite{HamausPoisson}:\footnote{We find
that our forecasts for $\sigma(\fnl), \sigma(\gnl), \sigma(\tnl)$ in this section are nearly unchanged if we use the Poisson approximation $E_{ij} \approx \delta_{ij}/{n_i}$
to Eq.~(\ref{eq:Eij}), except for a $\sim 10\%$ increase in the errors on the sample variance plateau.}
\ba
E_{ij} &=& \langle \epsilon_i \epsilon_j \rangle = \langle (\delta_i - b_i \delta) (\delta_j - b_j \delta) \rangle \nn \\
  &=& \langle \delta_i \delta_j \rangle - b_i \langle \delta_j \delta \rangle - b_j \langle \delta_i \delta \rangle + b_i b_j \langle \delta^2 \rangle  \nn \\
  &=& \frac{\delta_{ij}}{{n}_i} - b_i \frac{M_j}{\bar{\rho}} - b_j \frac{M_i}{\bar{\rho}} + b_i b_j \frac{\langle n M^2 \rangle}{\bar{\rho}^2}  \label{eq:Eij}
\ea
In the last line, we have taken the limit $k \rightarrow 0$ of the halo model predictions. Here we have defined
\be
\langle n M^2 \rangle = \int dM\, \frac{dn}{dM} M^2
\ee

Note that the two-halo contribution to $E_{ij}$ cancels entirely. The off-diagonal components have a contribution from the one-halo term, while the on-diagonal components are a sum of the usual Poisson-like term $1/n_i$ and one-halo contribution. It is possible to construct an estimator that weighs each halo bin optimally, which is going to be a compromise between reduction of Poisson shot noise (which would correspond to pure mass weighing) and cancellation of cosmic variance. As shown in \cite{Seljak_SV}, the Fisher Matrix formalism already includes these effects.

\begin{figure}[!ht]
\centerline{\includegraphics[width=12cm]{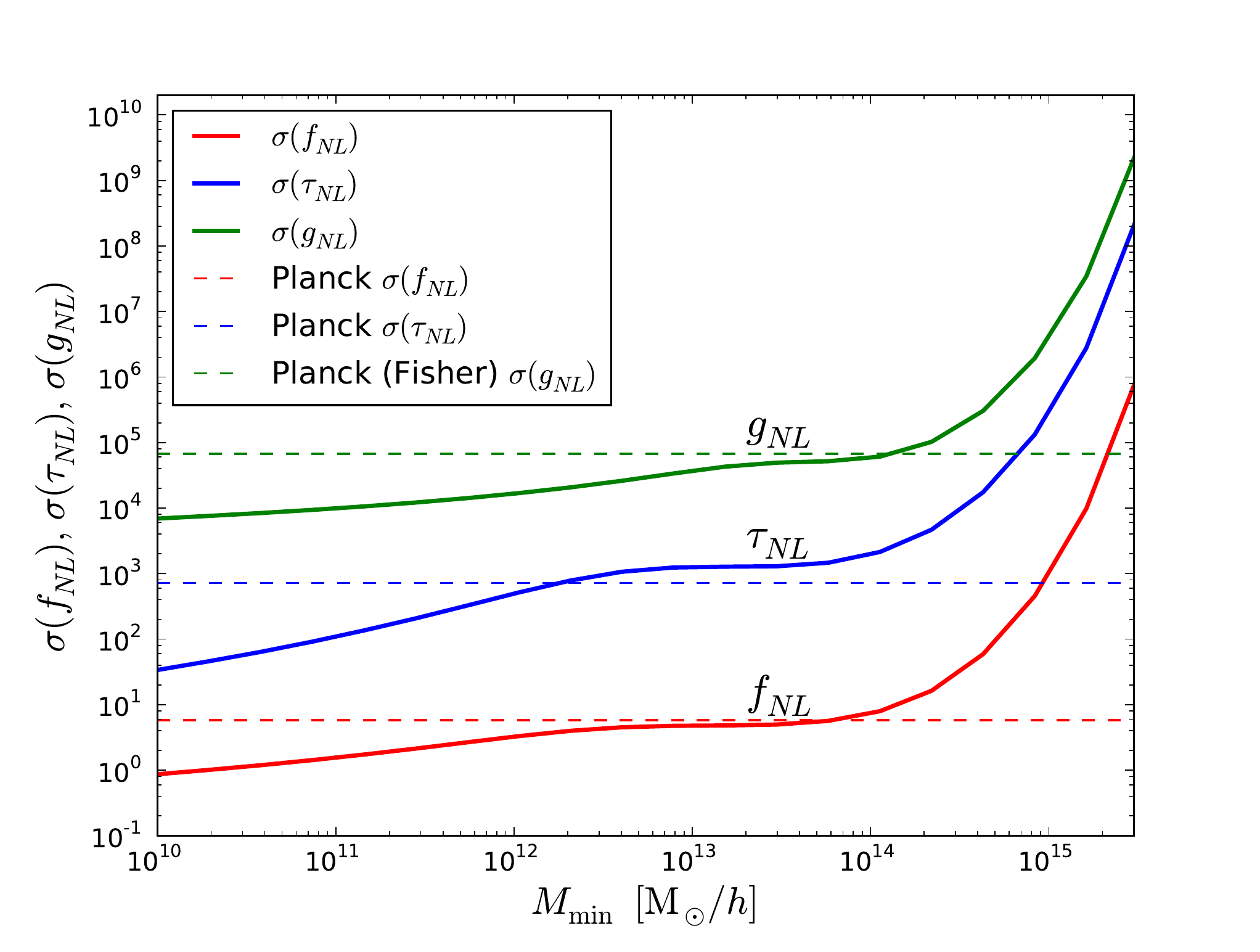}}
\caption{Statistical errors on $\fnl$ (bottom solid curve), $\tnl$ (middle solid curve) and $\gnl$ (top solid curve) in a multitracer analysis,
with varying $\Mmin$ and $N = 50$ mass bins equally spaced on a log scale. When forecasting a given parameter $\{\fnl,\tnl,\gnl\}$, the other two are set to zero. Here the volume is $V=25$ $h^{-3}$ Gpc$^3$, the redshift $z=0.7$ and $\kmax=0.1$ $h$ Mpc$^{-1}$. Note the `sample variance plateau' at $\Mmin \sim 3 \times 10^{13}$ $h^{-1} M_\odot $. The upper dashed line shows the Planck Fisher forecast $\sigma(\gnl) = 6.7 \times 10^4$ from \cite{Planckgnl}. The middle dashed line is the Planck $\sigma(\tnl) \approx 720$, obtained by fitting a Gaussian to the upper part of the $\tnl$ posterior for $L_{\rm max} = 50$ (Figure 19 of \cite{planck}).
\label{fig:mwfnltnl}
}
\end{figure}

In Figure~\ref{fig:mwfnltnl}, we show forecasted statistical errors $\sigma(\fnl)$, $\sigma(\tnl)$ and $\sigma(\gnl)$
from optimal weighting, for varying minimum halo mass $\Mmin$.
For high $\Mmin$ we are in the Poisson limited regime and the constraints from halo bias are not competitive with those from Planck.   As $\Mmin$ decreases, the statistical errors decrease rapidly, then plateau near $\Mmin \sim 5 \times 10^{13}$ $h^{-1} M_\odot $,
then decrease more slowly.

This ``sample variance plateau'' region can be interpreted as the range of $\Mmin$ where the tracer density is high enough to be sample variance limited, but not high enough that sample variance cancellation is effective. 
The sample variance plateau is important when thinking about survey optimization.
Once a survey is deep enough to reach the sample variance plateau, further improvements in survey depth do not
significantly improve constraints on primordial non-Gaussianity, unless the improvement is large enough ($\gtrsim 3$
magnitudes) to go past the plateau. Pushing to lower $\Mmin \lesssim 4 \times 10^{12} h^{-1} M_\odot $, cancellation of sample variance becomes effective with a moderate effect on $\fnl$ or $\gnl$, and a much larger one on $\tnl$, since for the latter case, most of the signal-to-noise comes from the very largest scales, which are the ones that are most affected by cosmic variance. 

From Figure~\ref{fig:mwfnltnl}, we see that a future generation with $V=25$ $h^{-3}$ Gpc$^3$ is competitive with Planck if resolving halos down to $\Mmin \sim 10^{14} h^{-1} M_\odot $. In order to significantly improve over Planck, either an increase in volume or a multi-tracer analysis with $\Mmin \lesssim 10^{13} h^{-1} M_\odot $ are needed.

\section{Separating $\fnl$, $\gnl$, $\tnl$}
\label{sec:separating_fgt}

So far, we have studied statistical errors on the parameters $\fnl, \gnl, \tnl$
individually, i.e.~we forecast the statistical error on each parameter assuming that the other
two parameters are zero\footnote{This assumption is not strictly consistent for
the case of $\fnl$, since $\tnl$ must satisfy the inequality $\tnl \ge (\frac{6}{5} \fnl)^2$
on general grounds.  However, we find that $\sigma(\tnl) \gg \sigma(\fnl)^2$ for all
forecasts considered in this paper, which implies that assuming
$\tnl=0$ when forecasting $\sigma(\fnl)$ is a good approximation to assuming the
`minimal' value $\tnl = (\frac{6}{5} \fnl)^2$}.
In this Section, we ask the question: to what extent can the parameters $\fnl,\gnl,\tnl$
be constrained jointly?

\subsection{Single tracer}
Considering the single-tracer case first, it is clear that $\fnl$ and $\gnl$ are
completely degenerate, since the clustering signature produced by $\fnl \ne 0$
is identical to the signature produced by $\gnl = (\beta_f/\beta_g) \fnl$.
On the other hand, there is some scope for separating $\fnl$ and $\tnl$ with
a single tracer, since the non-Gaussian bias has different scale dependence
in the two cases ($\fnl k^{-2} T(k)^{-1}$ versus $\tnl k^{-4} T(k)^{-2}$).
We can quantify this by using the Fisher matrix formalism to compute the
correlation coefficient
\be
\Corr(\fnl,\tnl) =  - \frac{F_{\fnl,\tnl}} {\sqrt{F_{\fnl} F_{\tnl}}}
\ee
where the minus sign appears because the covariance matrix is the inverse of the Fisher matrix.

\begin{figure}[!ht]
\centerline{\includegraphics[width=12cm]{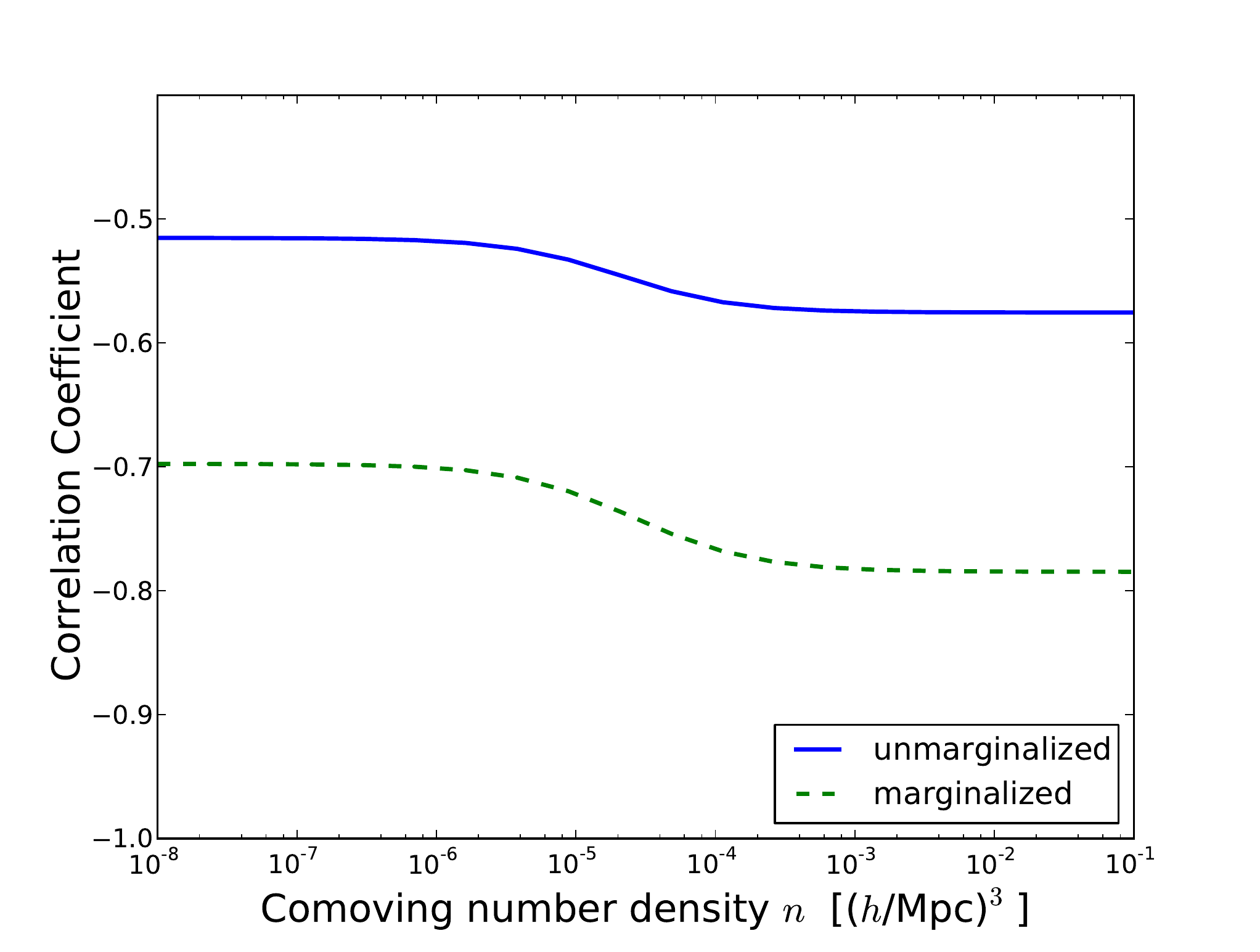}}
\caption{Single-tracer correlation coefficient between $\fnl$ and $\tnl$ in the unmarginalized case (top curve) and marginalizing over $b_g$ with a 20\% Gaussian prior on $1/\neff$ (bottom curve). The results are shown for our fiducial survey with $V=25$ $h^{-3}$ Gpc$^3$, $z=0.7$ and $b_g = 2.5$. }
\label{fig:corr_single} 
\end{figure}

An analytic calculation along the lines of Section \ref{SV_P_analytic} suggests that there should always be a moderate negative correlation between $\fnl$ and $\tnl$ in the single-tracer case. Figure \ref{fig:corr_single} shows the numerical results for our fiducial survey. Note that having to marginalize over $b_g$ and $1/\neff$ makes $\fnl$ and $\tnl$ more degenerate and harder to distinguish.

\subsection{Multiple tracer}
\begin{figure}[!ht]
\centerline{\includegraphics[width=12cm]{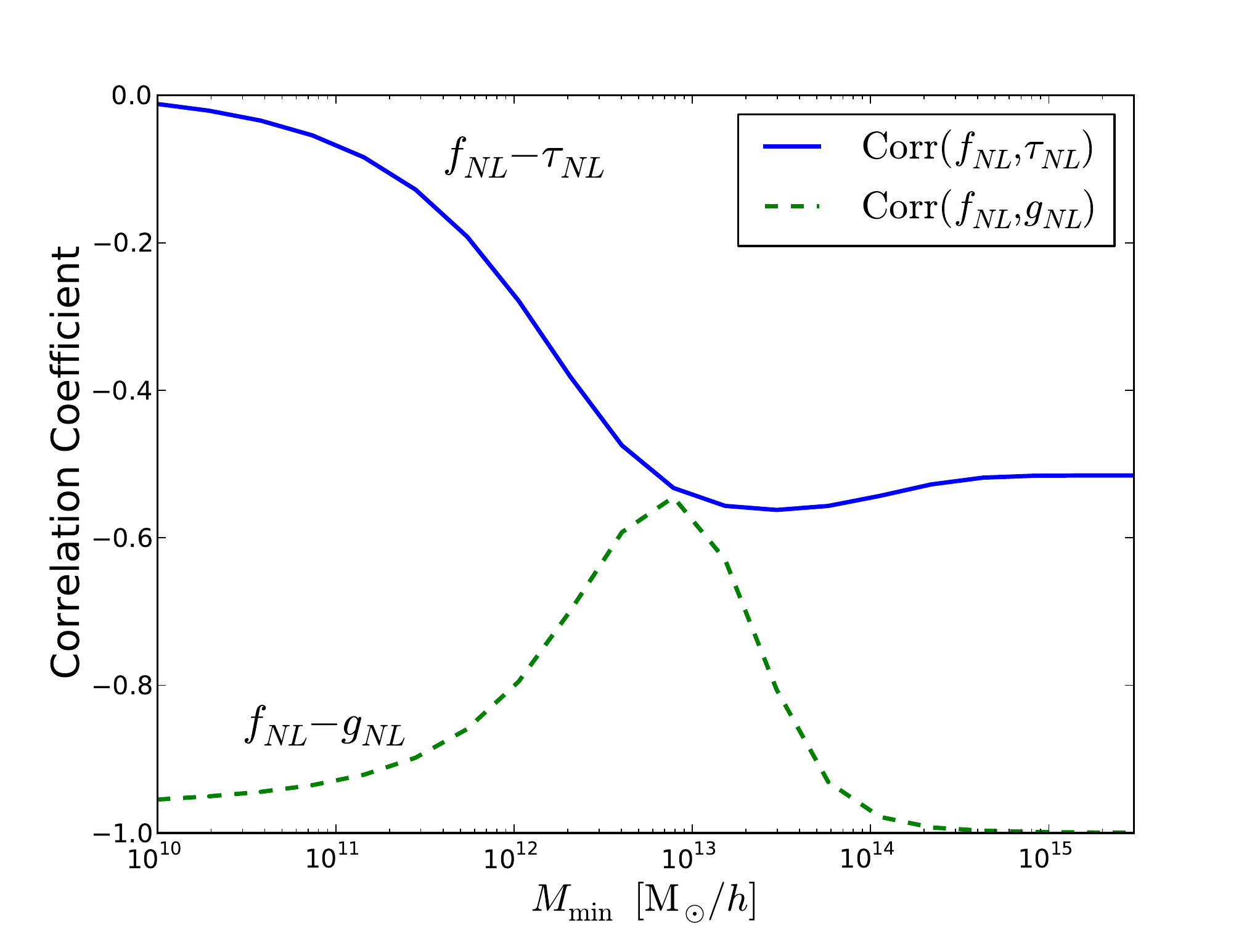} }
\caption{Multi-tracer correlation coefficients $\Corr(\fnl,\tnl)$ (top curve) and $\Corr(\fnl,\gnl)$ (bottom curve),
for varying minimum halo mass $\Mmin$.}
\label{fig:corr}
\end{figure}

The multi-tracer case is more interesting since $\fnl$ and $\gnl$ are no longer degenerate due to the different dependence of $\beta_f$ and $\beta_g$ on halo mass (or equivalently on Gaussian bias).
Following Section \ref{sec:multi_bin}, we assume perfect measurements of all halos above some minimum
mass $\Mmin$, and use the Fisher matrix formalism to compute the correlation coefficients $\Corr(\fnl,\tnl)$
and $\Corr(\fnl,\gnl)$. Numerical results are shown in Figure~\ref{fig:corr}.

Let's consider the $\fnl-\tnl$ case first. In the region with high $\Mmin$ the tracer density is low and we are deeply in the Poisson dominated regime, with correlation coefficient close to $-0.5$, in agreement with Figure \ref{fig:corr_single}. Decreasing $\Mmin$ allows more tracers to be included and the correlation becomes more negative, as expected from the previous discussion. As soon as $\Mmin$ reaches the sample variance plateau, $\fnl$ and $\tnl$ start to decorrelate, reaching nearly zero correlation at $\Mmin \sim 10^{10} h^{-1} M_\odot $.

Joint constraints on $\fnl, \tnl$ were also studied in~\cite{Biagetti}, who found
poor prospects for distinguishing the two, and generally weak constraints on $\tnl$,
if the stochastic bias from $\tnl$ is not included.  We therefore conclude that stochastic
bias is a very powerful observational probe of $\tnl$. 

In the $\fnl-\gnl$ case, the two are completely degenerate in the Poisson limit of high $\Mmin$ and are therefore observationally indistinguishable using halo bias. 
Close to the sample variance plateau they decorrelate partially, to become highly negatively correlated again in the region of sample variance cancellation.
We conclude that $\fnl,\gnl$ are not perfectly degenerate in a multi-tracer analysis, but are always significantly correlated (see also \cite{Roth} for a detailed discussion of the degeneracy between $\fnl$ and $\gnl$).

\section{Discussion and Conclusions}
\label{sec:discussion}
A detection of primordial non-Gaussianity would have very profound consequences for our understanding of the early Universe. Non-Gaussianity of the local type has been shown to leave an imprint on the large scale distribution of halos and galaxies in the form of a scale-dependent correction to the bias. Looking for this effect is one of the most promising ways to improve on the already tight bounds obtained by the Planck satellite.

In this work we have consider the effects of the scale-dependent bias on the power spectrum of halos and obtained forecasts applicable to upcoming Large Scale Structure surveys. 
Below we summarize our conclusions:

\begin{itemize}
\item If no mass information or other proxy for the bias of individual objects is available, a `single tracer' analysis is used. A survey volume $V = 25 h^{-3}$Gpc$^3$, median redshift $z = 0.7$ and mean bias $b_g = 2.5$, can achieve $\sigma(\fnl) = 6$, $\sigma(\gnl) = 10^5$ and $\sigma(\tnl) = 10^3$, if enough objects are resolved that the survey is sample variance limited.

\item The statistical error on $\fnl$ and $\gnl$ approximately scales like $V^{-2/3}$ and $V^{-1/2}$ in sample variance or Poisson domination regimes respectively. The error on $\tnl$ scales like $V^{-4/3}$ (sample variance domination) or $V^{-1}$ (Poisson domination).  In cases where the statistical error does not scale as $V^{-1/2}$, most of the statistical weight comes from the very largest scales in the survey.

\item When constraining primordial non-Gaussianity from large-scale structure,
it is always important to marginalize over Gaussian bias $b_g$ (and to a lesser extent, Poisson noise $1/\neff$) 
In particular, if $b_g$ is not marginalized in the sample variance dominated case, small increases in $\kmax$ can appear to produce a large improvement on statistical errors. This is not the case when proper bias marginalization is performed, since in this regime and for $k \gtrsim 10^{-1} h$ Mpc$^{-1}$, Gaussian bias and non-Gaussian corrections become nearly degenerate.

\item Neglecting redshift information in large-scale structure degrades
statistical errors on primordial non-Gaussianity by a factor close to 3.  However, redshift uncertainties of order $\Delta z \approx 0.1$ 
increase the errors by $\approx 1.4$ compared to the 
knowing the redshifts perfectly.  Therefore a next generation photometric survey will be able to extract most of the information.

\item A single-tracer sample variance limited survey with $V = 25 h^{-3}$Gpc$^3$ has a statistical power comparable to Planck. Improvement over CMB experiments would require either a larger volume or the use of multi-tracer techniques. If the mass or bias of individual objects is known, it is possible to combine different populations optimally in order to partially cancel sample variance and decrease the error. This mechanism becomes effective when resolving halos with $\Mmin \lesssim 10^{13} h^{-1} M_\odot $. If halos down to $\Mmin \sim 10^{11} h^{-1} M_\odot $ are resolved, we forecast $\sigma(\fnl) = 1.5$, $\sigma(\gnl) = 10^4$ and $\sigma(\tnl) = 100$, improving over Planck or a single-tracer analysis by a factor of 4 for $\fnl$ and nearly an order of magnitude for $\gnl$ and $\tnl$.

\item $\fnl$ and $\tnl$ can be distinguished even with a single tracer, due to the different scale dependence of the bias on large scales ($k^{-2}$ vs $k^{-4}$), but there is a significant correlation in the single tracer case.  They can be decorrelated by using a multi-tracer analysis and pushing to $\Mmin \lesssim 10^{13} h^{-1} M_\odot $. $\fnl$ and $\gnl$ are indistinguishable in the single-tracer case since the clustering signature produced by $\fnl \ne 0$ is identical to the that produced by $\gnl = (\beta_f/\beta_g) \fnl$. The multi-tracer case can make use of the fact that the non-Gaussian bias depends on the Gaussian bias in different ways to distinguish the two. However, the correlation coefficient is always close to $-1$.

\item Finally we briefly comment on survey optimization for primordial non-Gaussianity.  For most cases of practical interest the `sample variance plateau' makes it very hard to reach the regime in which sample variance cancellation becomes effective. So the most effective way of reducing the statistical errors is to increase the survey area (and hence the total volume), unless already resolving halos with masses at the lower end of the plateau ($\Mmin \sim 10^{13} h^{-1} M_\odot $ with our fiducial volume), in which case a deeper survey will also correspond to a significant improvement of statistical power.

\end{itemize}

\section*{Acknowledgements}
We thank Paolo Creminelli,  Marilena LoVerde, Emmanuel Schaan, Fabian Schmidt, Sarah Shandera, David Spergel, Michael Strauss and Matias Zaldarriaga for very helpful discussions and Pat McDonald for pointing out a typo to us.
SF is supported by NASA ATP grant NNX12AG72G, NSF grant AST1311756, and thanks Perimeter Institute for hospitality.
KMS was supported by an NSERC Discovery Grant.
Research at Perimeter Institute is supported by the Government of Canada
through Industry Canada and by the Province of Ontario through the Ministry of Research \& Innovation.

\bibliography{outline}

\appendix

 \begingroup\raggedright\endgroup

\end{document}